\documentclass[11pt,dvips]{article}
\usepackage{epsfig,times} 
\usepackage{picinpar}
\usepackage{wrapfig}
\usepackage{floatflt}
\usepackage{here}
\makeatletter
\renewcommand{\section}{\@startsection{section}{1}{0in}
	{0.4\baselineskip}{0.1\baselineskip}{\Large\bf}}
\renewcommand{\subsection}{\@startsection{subsection}{2}{0in}
	{0.25\baselineskip}{-\baselineskip}{\large\bf}}
\renewcommand{\subsubsection}{\@startsection{subsubsection}{3}{0in}
	{0.1\baselineskip}{-\baselineskip}{\normalsize\bf}}

\makeatother
\def\gtrsim{\mathrel{\hbox{\rlap{\hbox{\lower4pt\hbox{$\sim$}}}\hbox{$>$}}}} 
\begin{document}

%
%  Session and Paper Code:
\thispagestyle{myheadings}
%
%  ***INSTRUCTIONS:***  Replace `OG 9.9.9' in the command argument below
%			with your assigned session and paper code:
\markright{OG 2.1.23}
\begin{center}
%
%  ***INSTRUCTIONS:***  Replace `Instructions for Preparation of Manuscript'
%			with your paper's title:
{\LARGE \bf VHE gamma-ray observations of Markarian 501}
\end{center}

%  Author List:
\begin{center}
%
%  ***INSTRUCTIONS:***  Replace authors and addresses below with your own:
%
{\bf A.C. Breslin$^1$, 
I.H. Bond$^2$, 
S.M. Bradbury$^2$, 
J.H. Buckley$^3$, 
A.M. Burdett$^{4,2}$, 
M.J. Carson$^1$, 
D.A. Carter-Lewis$^5$,
M. Catanese$^5$, 
M.F. Cawley$^6$, 
S. Dunlea$^1$, 
M. D'Vali$^2$,
D.J. Fegan$^1$, 
S.J. Fegan$^4$, 
J.P. Finley$^7$, 
J.A. Gaidos$^7$,
T.A. Hall$^7$, 
A.M. Hillas$^2$, 
D. Horan$^1$, 
J. Kildea$^1$,
J. Knapp$^2$,
F. Krennrich$^5$, 
S. LeBohec$^5$, 
R.W. Lessard$^7$, 
C. Masterson$^1$,
B. McKernan$^1$, 
J. Quinn$^1$, 
H.J. Rose$^2$, 
F.W. Samuelson$^5$, 
G.H. Sembroski$^7$,
V.V. Vassiliev$^4$ and 
T.C.~Weekes$^4$}\\
{\it 
$^1$ Experimental Physics Dept., University College, 
Belfield, Dublin 4, Ireland\\ 
$^2$Dept. of Physics \& Astronomy, University of Leeds, 
Leeds, LS2 9JT, U.K.\\
$^3$Dept. of Physics, Washington University, St Louis, MO 63130, U.S.A.\\
$^4$Fred Lawrence Whipple Observatory, Harvard-Smithsonian CfA, 
Amado, AZ 85645-0097, U.S.A.\\
$^5$Dept. of Physics and Astronomy, Iowa State University, Ames, 
IA 50011, U.S.A.\\
$^6$Physics Dept., National University of Ireland, Maynooth, Ireland\\
$^7$Dept. of Physics, Purdue University, West Lafayette, IN 47907, U.S.A.}
\end{center}

%  Abstract:
\begin{center}
{\large \bf Abstract\\}
\end{center}
\vspace{-0.5ex}
%
%  ***INSTRUCTIONS:***  Replace text below with your own abstract:
%
Markarian 501, a nearby (z=0.033) X-ray selected BL Lacertae object,
is a well established source of Very High Energy (VHE, E$\gtrsim$300
GeV) gamma rays. Dramatic variability in its gamma-ray emission on
time-scales from years to as short as two hours has been
detected. Multiwavelength observations have also revealed evidence
that the VHE gamma-ray and hard X-ray fluxes may be correlated. Here
we present results of observations made with the Whipple
Collaboration's 10 m Atmospheric \v{C}erenkov Imaging Telescope during
1999 and discuss them in the context of observations made on Markarian
501 during the period from 1996-1998.

%

%  Leave this line skip in place:
\vspace{1ex}

%
%  Manuscript text:
%
%  ***INSTRUCTIONS:***  Delete the next few pages of text and enter your own.  There will
%			be a warning, `STOP DELETING TEXT!!', just before the References
%			section so that the standardized Reference heading will not be
%			accidently erased.  Within the text below is an example is given
%			of a figure placement (using `picinpar').
\section{Introduction:}
\label{intro.sec}
In 1995, the Whipple Collaboration discovered Very High Energy (VHE,
E$\gtrsim$300 GeV) gamma rays from Markarian 501 (Mkn501) (Quinn et
al., 1996) using the 10~m Atmospheric Imaging \v{C}erenkov Telescope
(IA\v{C}T). This detection has since been confirmed by several other
IA\v{C}T experiments (e.g. see Protheroe et al., 1997). Markarian 501
is one of four active galactic nuclei (AGN) detected at VHE energies,
the others being Markarian 421 (Mkn421) (Punch et al., 1992),
1ES2344+514 (Catanese et al., 1998) and PKS2155-304 (Chadwick et al.,
1999). These objects all exhibit optical properties characteristic of
the BL Lacertae (BL Lac) class of AGN and, due to the extension of
their synchrotron spectra into the X-ray band, are classified as X-ray
Selected BL Lacs (XBLs).  Markarian 501 has been detected at the
4$\sigma$ level by the Energetic Gamma-Ray Experiment Telescope
(EGRET) on board the Compton Gamma-Ray Observatory (Kataoka et al.,
1999).

Since 1995 the Whipple Collaboration has extensively monitored the
gamma-ray emission from Markarian 501. The VHE flux level has been
observed to vary dramatically on a range of different timescales
(Quinn et al., 1999). The mean flux level has varied from $\sim$0.1 to
$\sim$5 times that of the Crab Nebula. Significant variability has
been observed on a range of timescales from years to hours.
Extreme variability has also been observed in Mkn421 (e.g. Gaidos et
al., 1996). However, a major difference in the flux characteristics of
the two objects is that Mkn501 appears to have a base emission level,
whereas any such base level in the case of Mkn421 lies below the
minimum sensitivity level of the Whipple 10~m telescope.

There is as yet no consensus on the dominant gamma-ray production
mechanisms in BL Lacs. The preferred model is that emission is
associated with the accretion of matter onto a supermassive black
hole, the generation of magnetic fields and the production of jets of
electrons, although the precise manner in which the beam is generated
and collimated is unclear (Buckley et al., 1997). Jet models are
consistent with evidence that the emission is beamed. The observed
correlation of the variability of TeV gamma rays and hard X-rays
implies that both are produced by non-thermal electrons. Furthermore,
if the synchrotron emission at energies $\gtrsim$ 100 keV detected by
OSSE is also attributed to the same source, then there is a test for
beaming in XBLs (Catanese at al., 1997). Observations of emission from
Mkn501 up to 7.5 TeV (Samuelson et al., 1999) show clear evidence of
beaming, with a Doppler boost factor of 1.5-2.0. The short timescales
of correlated variability put more restrictive constraints on a proton
beam, which is assumed to lose energy mainly by synchrotron
radiation. The necessary magnetic field of 30-90 G for a boost factor
of the order of 10 are not excluded and would be consistent with the
variability of Mkn421 (Mannheim et al., 1998).

\section{Observations and Analysis:}
\label{obs.sec}
Observations were made with the Whipple IA\v{C}T, located on
Mt.\ Hopkins in southern Arizona. The 10~m optical reflector focuses
 \v{C}erenkov light from gamma-ray and cosmic-ray initiated
air-showers onto a high resolution camera mounted in the focal
plane. Subsequent off-line analysis of images identify candidate
gamma-ray events. The camera, which comprises a close-packed array of
331 photomultiplier tubes (PMTs), has a field of view of
4.8$^\circ$. The telescope is triggered when two out of the 331 PMTs
register a signal which exceeds some pre-set threshold. This is
typically achieved by summing the outputs of a discriminator on each
channel and putting the summed signal through another
discriminator. However, a disadvantage of this method is that random
two-pixel events, caused by fluctuations in the night sky background,
can trigger the system.  The rate at which such unwanted triggers
occur is a major factor in determining the operating threshold of the
telescope.

A pattern selection trigger (PST), which can be pre-programmed with
acceptable PMT trigger combinations, has been developed (Bradbury et
al., 1999). This permits a decrease in the operational energy
threshold of the telescope. During the 1998/1999 observing season the
PST was initially tested and subsequently used for routine data
taking. As a result, the Markarian 501 dataset consists of two
components, one taken at a slightly higher threshold than the
other. Based on observations of the Crab Nebula, we estimate that the
energy thresholds are 500~GeV and 400~GeV respectively. The same
selection criteria have been used for both datasets and the effective
energy threshold with the PST is lower than 400 GeV with an optimised
analysis.

For the analysis presented here, only data taken under good weather
conditions and with elevations greater than 50$^\circ$ have been
used. Our data set consists of 4.5 hours taken at an energy
threshold of $\sim$500~GeV and 15.5 hours taken at the $\sim$400~GeV
threshold, spanning a total of 20 nights in the period February 1999
to April 1999.  The integral fluxes obtained with the
different thresholds have been normalised by expressing them as
fractions of the measured Crab Nebula flux (CF)  at the same threshold.
This assumes that the slopes of the VHE energy spectra of Mkn 501 and
the Crab Nebula are similar, which is true only to a first
approximation (Samuelson et al., 1998, Hillas et al., 1998).

\section{Results:}
\label{res.sec}
Figure 1 shows the averaged gamma-ray rate expressed in terms of the
Crab Nebula rate over three month's of observation during 1999.  The
mean rate of gamma rays from Markarian 501 is 0.36$\pm$0.025 CF for
the season. This rate is comparable to the measured rate in 1998 and
significantly higher than that in 1995 or 1996 (Quinn et al.,
1999). There is strong evidence for variability (figure 1) with the
rate on the first night (MJD 51224) of observation being considerably
higher than for the rest of the observations. A test for constant flux
level gives a $\chi^2/DOF$ of 107/19, with a chance probability
$\sim$10$^{-14}$. Omitting this first night still suggests variability
with a chance probability of $2\times10^{-4}$. There is also
evidence for variability in the monthly averages when the first data
point is omitted, with average rates of $0.60\pm 0.065$, $0.33\pm
0.045$ and $0.26\pm 0.034$ CF respectively for February, March and
April, giving a probability for constant flux $\sim2\times10^{-5}$. 
Data from MJD 51228
were excluded because of extremely poor weather conditions, although
there was a strong signal of approximately 3 CF on
that night.

Previous multiwavelength observations indicated that the VHE gamma-ray
and X-ray data may be correlated (e.g. Catanese at al., 1997). We have
plotted the average weekly rates for data obtained with the Whipple
10~m IA\v{C}T in figure 2 (top) and the available X-ray data from the
RXTE All Sky Monitor (ASM) in figure 2 (bottom) for the period from
1996 to 1999. There is evidence for some correlation in long-term
trends in both wavebands. It is difficult to determine correlations on
shorter time-scales due to poor sampling in the VHE band and low
statistics in the ASM data.  The large flare detected at the Whipple
Observatory on MJD 51224 seems to have preceded a slightly elevated
state in the X-ray emission, which lasted for a few weeks.

%\begin{figwindow}[1,r,%
%{\mbox{\epsfig{file=icrc_rate_n.eps,width=4.5in}}},%
%{Average nightly rates expressed in terms of the Crab Nebula flux (CF)
%measured with the Whipple 10~m IA\v{C}T in 1999}]
%
%\end{figwindow}

\begin{figure}
\begin{center}
\epsfig{file=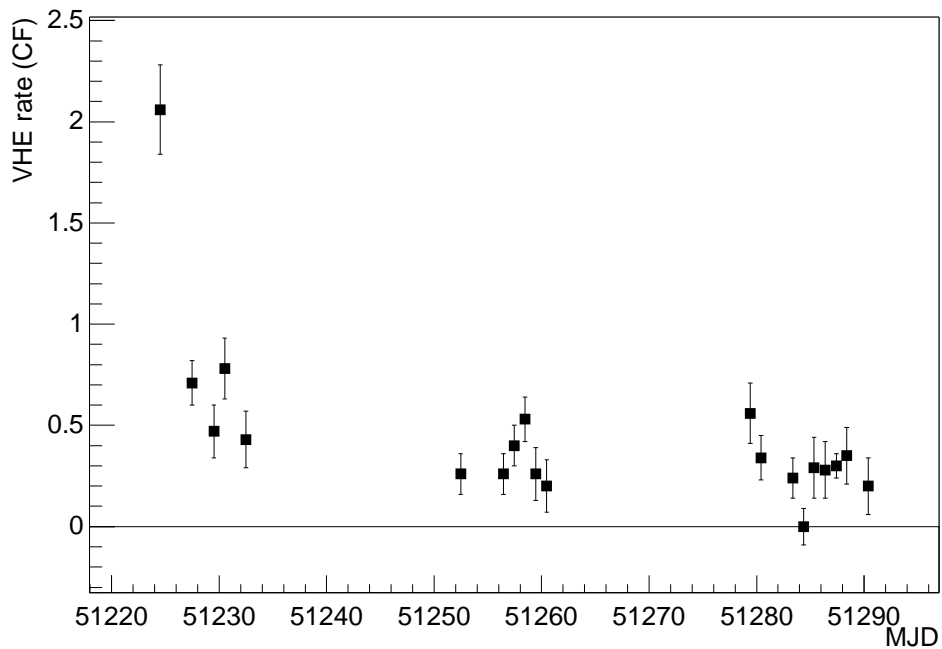,width=5in}
\caption{Average nightly rates expressed in terms of the Crab Nebula flux (CF)
measured with the Whipple 10~m IA\v{C}T in 1999}
\end{center}
\end{figure}

%\begin{figure}
%\begin{center}
%\epsfig{file=icrc_fig1.eps,width=17cm}
%\caption{The flux variability of Markarian 501 in VHE gamma-rays
%(top) and X-rays from the RXTE ASM (bottom). While there is evidence
%for variability in both data-sets there is not strong evidence
%for correlated variability.}
%\end{center}
%\end{figure}

%\begin{figure}
%\epsfig{file=icrc_mw.eps,width=4.5in}
%\end{figure}

%\label{Average weekly rates from Markarian 501 from the Whipple IA\v{C}T (top) and the RXTE ASM (bottom) for the period 1996-1999}

\section{Conclusions:}
\label{con.sec}
%\begin{figwindow}[1,r,%
%{\mbox{\epsfig{file=icrc_mw.eps,width=4.5in}}},%
%{Average weekly rates for Markarian 501 from the Whipple
%IA\v{C}T (top) and the RXTE ASM (bottom) for the period 1996-1999}]
% 
The results provide clear evidence for continued variability in Mkn501
in 1999. There is strong evidence for day-scale variability, with the
emission on the first night of observation (MJD 51224) considerably
higher than on any other night during the rest of the observation
period.  Even omitting this night from the dataset there is 
still some evidence for variability on timescales of days and months.
The average emission level is significantly below that of
1997, but comparable to the level of 1998.  Once again, there is
evidence for a baseline emission level which possibly varies on
time-scales of several months or longer.

Comparison of the ASM X-ray and Whipple VHE data from 1996 to 1999
show similarities in the long term trends but not in individual
flaring episodes. More detailed studies of correlations between the
multiwavelength light-curves are under way and will be reported at
this conference.  Also to be reported are the results of more
intensive multiwavelength campaigns undertaken during Spring 1999.
%\end{figwindow}

%Variations on time-scales of weeks or months were not detected.

%\begin{figwindow}[1,r,%
%{\mbox{\epsfig{file=icrc_mw.eps,width=4.5in}}},%
%{Average weekly rates from Markarian 501 from the Whipple
%IA\v{C}T (top) and the RXTE ASM (bottom) for the period 1996-1999}]
%
%\end{figwindow}

\section{Acknowledgements:}
We acknowledge the technical assistance of K.\ Harris and E.\
Roache. This research is supported by the U.S. Department of Energy,
by PPARC (U.K.) and Enterprise Ireland. The X-ray data in this work
are quicklook results presented by the ASM/RXTE team.
\vspace{1ex}

\clearpage

\begin{figure}[H]
\begin{center}
\epsfig{file=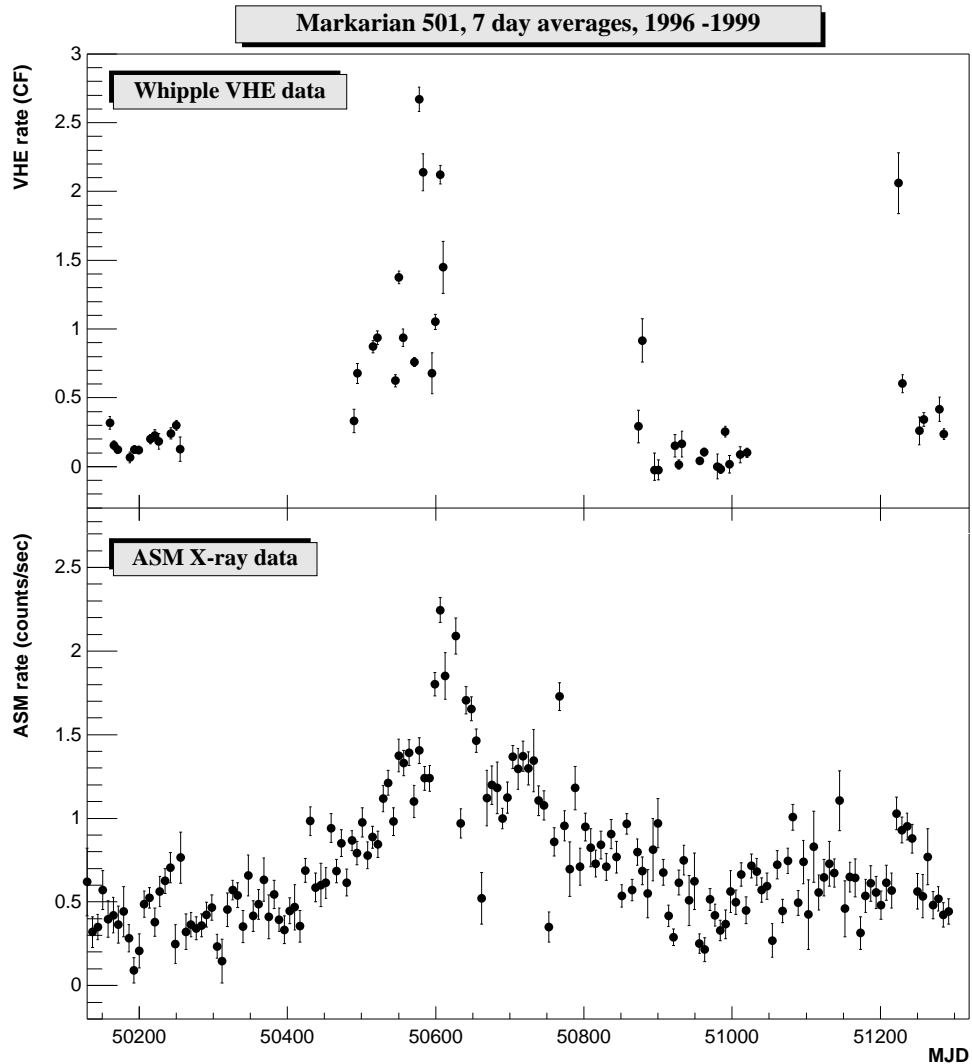,width=5.2in}
\caption{Average weekly rates for Markarian 501 from the Whipple
IA\v{C}T (top) and the RXTE ASM (bottom) for the period 1996-1999}
\end{center}
\end{figure}

\begin{center}
\vspace*{-4ex}
{\Large\bf References}
\end{center}
%
%  ***INSTRUCTIONS:***  Enter your references alphabetically following the format
%			of the example citations below.
\vspace*{-1.5ex}
Bradbury, S.M. et al. 1999, Proc. 26th ICRC (Salt Lake City, 1999), OG
4.3.21\\   
Catanese, M. et al., 1997, ApJ, 487, 143\\ 
Catanese, M. et al., 1998, ApJ, 501, 616\\ 
Chadwick, P.M. et al., 1999, ApJ, 515,161\\ 
Gaidos, J.A. et al., 1996, Nature, 383, 319\\ 
Hillas, A.M. et al., 1998, ApJ, 503, 744\\ 
Kataoka, J. et al., 1999, ApJ, 514, 138\\
Mannheim, K. et al., 1998, Science, 279, 684\\
Punch, M. et al., 1992, Nature, 358, 477\\ 
Protheroe, R.J. et al., 1998, Proc. 25th ICRC (Durban, 1997), 8, 317\\ 
Quinn, J. et al., 1996, ApJ, 456, L83\\ 
Quinn, J. et al., 1999, ApJ, in press\\
Samuelson, F.W. et al., 1998 ApJ, 501, L17
\end{document}